\documentclass[pre,twocolumn,floatfix,superscriptaddress,longbibliography,nofootinbib]{revtex4-1}
\RequirePackage[sort&compress]{natbib}

\usepackage{color}

\usepackage{amssymb,amsfonts,amsmath}
\usepackage{bm}
\usepackage[pdftex]{graphicx}
\usepackage{xcolor}
\usepackage{subfigure}
\usepackage{comment}

\usepackage{hyperref}
\hypersetup{
    colorlinks=true,
    linkcolor=blue,
    filecolor=magenta,      
    urlcolor=blue,
}
 
\definecolor{darkGray}{RGB}{153,153,153}
\definecolor{darkBlue}{RGB}{37,113,161}
\definecolor{darkGreen}{RGB}{113,161,37}
\definecolor{darkRed}{RGB}{186,21,24}

\newcommand{\bra}[1]{\mbox{$\langle #1|$}}
\newcommand{\ket}[1]{\mbox{$|#1\rangle$}}

\DeclareMathOperator{\Tr}{Tr}

\begin{document}

\title{Spectral entropies as information-theoretic tools for complex network comparison}

\author{Manlio De Domenico}
\affiliation{Departament d'Enginyeria Inform\`{a}tica i Matem\`{a}tiques, Universitat Rovira i Virgili, 43007 Tarragona, Spain}

 \author{Jacob Biamonte}
\affiliation{Quantum Complexity Science Initiative \\ Department of Physics, University of Malta, MSD 2080 Malta}
\affiliation{Institute for Quantum Computing \\ University of Waterloo, Waterloo, N2L 3G1 Ontario, Canada}

\begin{abstract}

{\bf Any physical system can be viewed from the perspective that information is implicitly represented in its state.  However, the quantification of this information when it comes to complex networks has remained largely elusive. In this work, we use techniques inspired by quantum statistical mechanics to define an entropy measure for complex networks and to develop a set of information-theoretic tools, based on network spectral properties, such as R\'{e}nyi $q-$entropy, generalized Kullback-Leibler and Jensen-Shannon divergences, the latter allowing us to define a natural distance measure between complex networks.
First we show that by minimizing the Kullback-Leibler divergence between an observed network and a parametric network model, inference of model parameter(s) by means of maximum-likelihood estimation can be achieved and model selection can be performed appropriate  information criteria. Second, we show that the information-theoretic metric quantifies the distance between pairs of networks and we can use it, for instance, to cluster the layers of a multilayer system. By applying this framework to networks corresponding to sites of the human microbiome, we perform hierarchical cluster analysis and recover with high accuracy existing community-based associations. Our results imply that spectral based statistical inference in complex networks results in demonstrably superior performance as well as a conceptual backbone, filling a gap towards a network information theory.
}
\end{abstract}

\maketitle

\section{Introduction} 

Shannon's entropy~\cite{shannon48} and information-theoretic derived measures have been successfully applied in a range of disciplines, from revealing time scale dependence in neural coding~\cite{borst1999information,strong1998entropy} to quantifying quantum information~\cite{holevo1973bounds,bennett1998quantum,vedral2002role} and the complexity of genetic sequences~\cite{mantegna1994linguistic,bernaola2000finding} or to unravel the mesoscale organization of interconnected systems~\cite{rosvall2008maps,esquivel2011compression,rosvall2014memory,dedomenico2015modular}, to cite just a few emblematic achievements.  However, when it comes to complex networks, an appropriate definition of entropy has remained elusive---applicability being often limited to the probability distribution of some network descriptor (such as the normalized distribution of node degrees). 

Complex network theory dates to the discovery of several fundamental features~\cite{WS98,barabasi1999emergence} and complex networks have been widely used to model and better understand the organization of complex systems and their dynamics~\cite{guimera2005functional,palla2005uncovering,song2005self,colizza2006detecting,boguna2009navigability,vespignani2012modelling}, the controllability of their constituents~\cite{liu2011controllability} and their resilience to structural and dynamical perturbations~\cite{albert2000error,callaway2000network,buldyrev2010catastrophic,gao2012networks,radicchi2013abrupt,dedomenico2014navigability}. A recent and ongoing drive in complex networks is to merge ideas with quantum information science, including entangled networks for communication~\cite{acin2007entanglement,cuquet2009entanglement,perseguers2010quantum}, developing a theory of node centrality in quantum walks on graphs~\cite{faccin2013degree} and in the detection of community structures formed in quantum systems~\cite{faccin2014community}. 

In the same spirit as the Church--Turing--Deutsch principle \cite{Deutsch97}, like all physical systems it is possible to view complex networks in terms of information processing, in which information changes in time from an input to an output state of a system. This then necessitates a method to quantify this information, its time-dependence in terms of storage and transfer of information, ideally between levels of a multilayer system, which can further be imperfect due to randomness or noise. A similar challenge was addressed by quantum theorists beginning decades ago when faced with quantifying informatic properties of quantum states \cite{RevModPhys.81.865, vedral2002role, holevo1973bounds, bennett1998quantum,wilde, wehrl1978general, araki1970}. Indeed, the modern theory of quantum information, as the name suggests, is fundamentally built upon the quantification of physical information, entropic based quantifications of non-locality, entanglement \cite{RevModPhys.81.865} and the inherent complexity of the quantum model~\cite{wilde}. These decades of research have placed entropic measures central to the modern theory, leading to quantum generalizations of Shannon's classical information theory~\cite{RevModPhys.81.865, vedral2002role, holevo1973bounds, bennett1998quantum,wilde, wehrl1978general, araki1970}. However, for a variety of reasons, these results can't be applied directly to complex networks, although classical information-theoretic tools have been successfully adopted to solve specific problems~\cite{rosvall2007information,rosvall2008maps,dehmer2008information,ludtke2008information,bianconi2009assessing,eagle2010network,rosvall2010mapping,johnson2010entropic,anand2011shannon,kovacs2015unified}.

Here we address a similarly motivated challenge: inspired by how entropy is calculated in quantum systems, we define an interconnectivity-based density matrix to calculate the von Neumann entropy of a network. We analytically prove that our definition satisfies the desired additivity properties similar to quantum thermodynamic entropy---though the proof of this differs tremendously from the quantum case---which opens the door to avoid the shortcomings, recently pointed out in \cite{dedomenico2015structural}, imposed by sub-additivity failing as found in past approaches.

We exploit this entropy to develop a set of information-theoretic tools which apply to complex networks, such as R\'{e}nyi q-entropy, generalized Kullback-Leibler and Jensen-Shannon divergences, thus providing a backbone to an information-theoretic approach to network science. Our framework, based on spectral properties of networks, allows one to probe contemporary problems faced in complex network science. For instance, we show that Kullback-Leibler minimization can be used to infer the parameter(s) of a model aimed to fit an observed network. By exploiting results of classical information theory, we are able to introduce the spectral counterpart of likelihood maximization and use it in practical applications to fit network models and perform model selection based on appropriate Akaike and Bayesian information criteria, as well as minimum description length. The strength of our approach relies on the fact that it uses the network as a whole, instead of a subset of network's descriptors, to attack parameter inference and model selection problems.

Another important byproduct of the proposed information theoretic framework is the possibility to quantify the distance between complex networks. This problem is of high interest in recent applications concerning multilayer systems, such as time-varying and multiplex networks~\cite{kivela2013multilayer,boccaletti2014structure,dedomenico2016physics}. This type of systems consist of nodes replicated across several networks that exhibit different types of relationships or connectivity~\cite{dedomenico2013mathematical}. The possibility to compare layers from an information-theoretic point of view has been recently explored to aggregate them in order to reduce the structure and the complexity of biological, transportation and social multiplex networks~\cite{dedomenico2015structural}. To numerically probe our method, we cluster the layer of an empirical system and compare it against the existing classification.

\section{von Neumann entropy of a complex network}

In quantum mechanics, probability distributions are encoded by density matrices. A density matrix $\boldsymbol{\rho}$ is a Hermitian and positive semi-definite matrix, with trace equal to unity which is used to represent both mixed and pure quantum states. A system is in a pure state $\ket{\psi}$, if and only if the bound $\Tr{\boldsymbol{\rho}^{2}}\leq 1$ is saturated. 
The density matrix admits a spectral decomposition as
\begin{eqnarray}
\boldsymbol{\rho}=\sum_{i=1}^{N}\lambda_{i}\ket{\phi_{i}}\bra{\phi_{i}}
\end{eqnarray}
for an orthonormal basis $\{\ket{\phi_{i}}\}$, where $\lambda_{i}$ are non-negative eigenvalues which sum up to 1.

The density matrix allows to define the von Neumann entropy by
\begin{eqnarray}\label{eq:vne}
S(\boldsymbol{\rho})&=&-\Tr{(\boldsymbol{\rho} \log_{2} \boldsymbol{\rho})}=-\sum_{i=1}^{N}\lambda_{i}\log_{2}\lambda_{i},
\end{eqnarray}
i.e.~it is equal to the Shannon entropy of the eigenvalues of the density matrix, where by convention $0 \log_2 0 := 0$.

In this work we are not limited in having a quantum setup where the matrix $\boldsymbol{\rho}$ can be called a ``density matrix'' in the physical sense, but instead build on this idea to define a matrix from a network that satisfies the same mathematical properties of a density matrix.

First, such a density matrix should be positive definite and symmetric, therefore $\boldsymbol{\rho}=Z^{-1}e^{-\beta \mathbf{H}}$ -- with $\mathbf{H}$ a symmetric matrix with non-negative eigenvalues, $Z$ and $\beta$ real numbers -- is a suitable candidate. Second, the eigenvalues of $\boldsymbol{\rho}$ must sum to unity, thus imposing the constraint $Z=\Tr{e^{-\beta \mathbf{H}}}$. Here, we use the following density matrix, defined by
\begin{eqnarray}
\label{eq:def-rho}
\boldsymbol{\rho} = \frac{e^{-\beta \mathbf{L}}}{Z},\quad Z=\Tr{e^{-\beta \mathbf{L}}},
\end{eqnarray}
where $\mathbf{L}=\mathbf{D}-\mathbf{A}$ denotes the combinatorial graph Laplacian, being $\mathbf{A}$ the adjacency matrix of the network and $\mathbf{D}$ the diagonal matrix of node degrees. Eq.~(\ref{eq:def-rho}) resembles the Gibbs state of a quantum system where the Hamiltonian equals the Laplacian matrix, i.e. $\mathbf{H}=\mathbf{L}$, where $\beta$ is a scale parameter. 

It is worth remarking that this functional form appears in many problems of physical interest. For instance, if $\beta$ parameterizes time, setting $Z=1$ and $\mathbf{H}=\mathbf{L}$, i.e.~the Laplacian matrix,  then $e^{-\mathbf{L}t}$ is the evolution matrix governing purely diffusive dynamics and its trace $\sum\limits_{i=1}^{N}e^{-\lambda_{i}t}$ is $N$ times the average return probability, i.e.~the probability that a walker starting at any node $i$ will return to its origin at time $t$. Another case of interest is when $\beta$ parameterizes temperature of a quantum system with Hamiltonian $\mathbf{H}$ and $Z=\Tr{e^{-\beta\mathbf{H}}}$, so that $\boldsymbol{\rho}$ would provide the Gibbs state of the system in equilibrium at finite temperature. Although we are interested mainly in non-negative values of $\beta$, it is interesting to show that when $\beta=-1$, $Z=1$ and $\mathbf{H}=\mathbf{A}$ we recover the well-known communicability matrix introduced by Estrada~\cite{estrada2008communicability,estrada2008communicabilitymulti,estrada2010network,grindrod2011communicability,ding2014dynamics,estrada2014communicability} (see Ref.~\cite{estrada2012physics} for a thorough review) or a normalized version of it, if $Z=\Tr{e^{\mathbf{A}}}$.

In Ref.~\cite{braunstein2006laplacian}, a matrix is built in such a way that its mathematical properties satisfy the requirements of a density matrix. For an undirected complex network, $\mathbf{L}$ is symmetric and positive semi-definite but its eigenvalues do not sum to unity. However, the eigenvalues of the proposed matrix, defined by $\boldsymbol{\rho}_{bgs}=\mathbf{L}/\Tr{\mathbf{L}}$, evidently do, therefore $\boldsymbol{\rho}_{bgs}$ is a mathematically suitable density matrix~\cite{braunstein2006laplacian,anand2011shannon} and hence, Eq.~\eqref{eq:vne} can in principle be applied. This approach has been recently generalized to the case of multilayer systems~\cite{dedomenico2013mathematical}, composite networks where units exhibits different types of relationships that are generally modeled as different layers and used to reduce their structure~\cite{dedomenico2015structural}.

However, it has been recently found that the von Neumann entropy calculated from the rescaled Laplacian $\boldsymbol{\rho}_{bgs}$ does not satisfy the sub-additivity property in some critical circumstances~\cite{dedomenico2015structural}. For instance, let $\boldsymbol{\rho}_{bgs}$ be the rescaled Laplacian matrix of a network with $N$ nodes and $|E|\gg1$ edges, let $\boldsymbol{\sigma}_{bgs}$ be the rescaled Laplacian matrix of a network with $N$ nodes and just one (undirected) edge and let $\boldsymbol{\tau}_{bgs}$ be the rescaled Laplacian of the network obtained by summing up, entry wise, the corresponding adjacency matrices of the previous two networks. Since $S(\boldsymbol{\sigma}_{bgs})=0$, the sub-additivity property $S(\boldsymbol{\tau}_{bgs})\leq S(\boldsymbol{\rho}_{bgs}) + S(\boldsymbol{\sigma}_{bgs})$ is not always satisfied because, as we will see later, $S(\boldsymbol{\tau}_{bgs})\geq S(\boldsymbol{\rho}_{bgs})$ very often. 

\begin{figure}[!t]
\centering
\includegraphics[width=.5\textwidth]{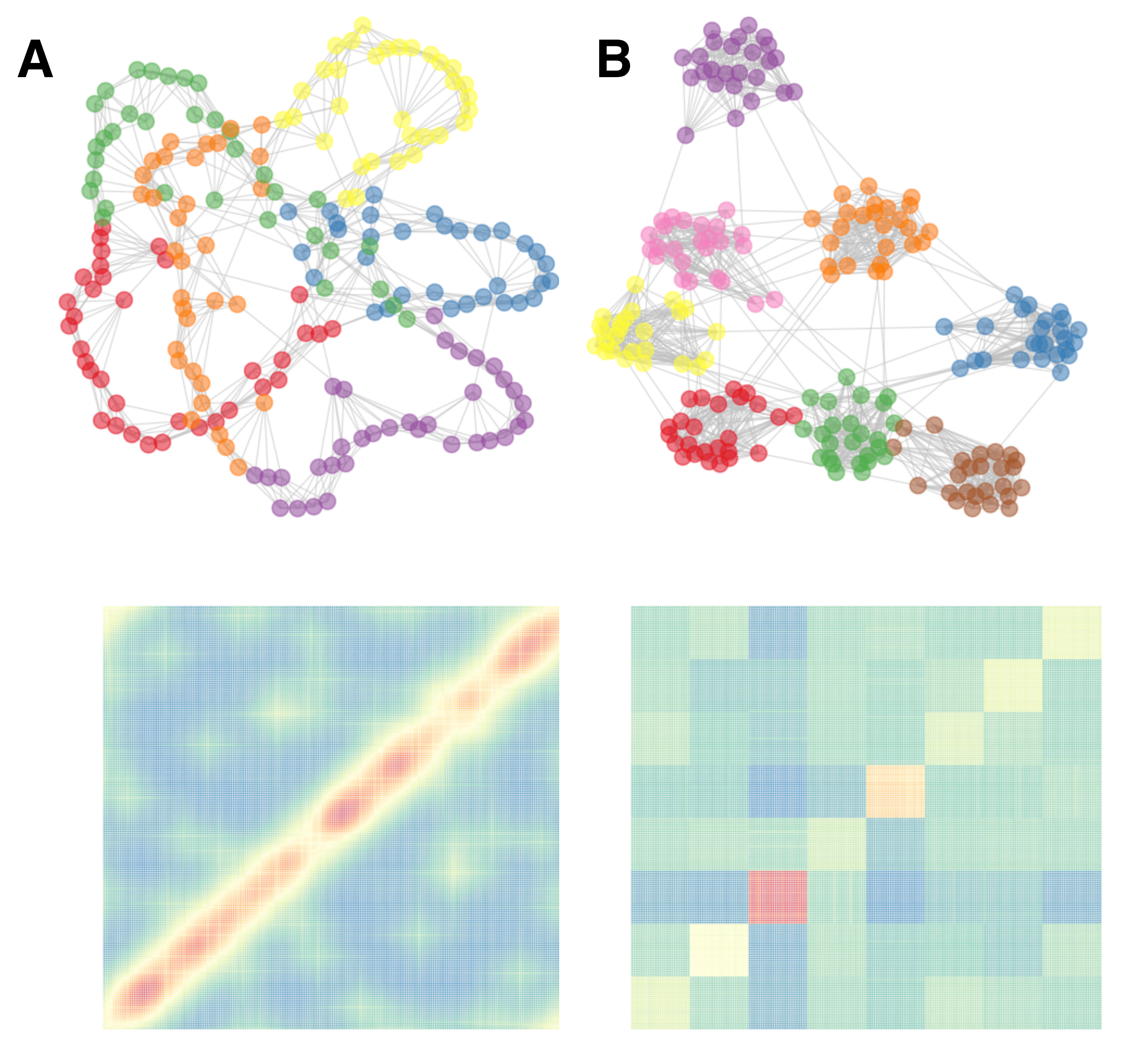}
\caption{\textbf{Density matrix of a complex network.} Network (top) and corresponding density matrix (bottom) from (A) a stochastic block-model and (B) Watts-Strogats networks, for $\beta=3.16$. Colors in the networks indicate nodes belonging to the same community, whereas colors in the density matrices indicate the magnitude of the corresponding entry. Networks with $N=200$ nodes are shown.}\label{fig:density-matrices}
\end{figure}

While this peculiar behavior can in fact be exploited in certain situations~\cite{dedomenico2015structural}, one is interested in preserving sub-additivity, because in general it is expected that the entropy of a composite system $A+B$ is equal or smaller than the sum of the entropy of $A$ and $B$. In the case of complex networks, it seems intuitively reasonable but has remained to be proven---see Appendix~\ref{app:subadd}---that the aggregation of two graphs is expected to have lower entropy than the sum of the entropy of each graph separately. The entropy of the density matrix introduced to complex networks in this work ---that we name spectral entropy---presents some interesting features that will be discussed later. A visualization of the density matrix corresponding to different network models is shown in Fig.\,\ref{fig:density-matrices}.

\section{von Neumann entropy of standard network models}

For simplicity, in the following we will use $\boldsymbol{\rho}=\boldsymbol{\rho}_{G}$ where there will be no ambiguity about which density matrix is under consideration. Indeed, we will also use the notation $S(G)=S(\boldsymbol{\rho}_{G})$ to indicate the entropy of network $G$.

From the eigen-decomposition of the Laplacian matrix $\mathbf{L}=\mathbf{Q}\mathbf{\Lambda}\mathbf{Q}^{-1}$, where $\mathbf{\Lambda}$ is the diagonal matrix of Laplacian's eigenvalues, it is straightforward to show that
\begin{eqnarray}
Z=\sum_{i=1}^{N}e^{-\beta \lambda_{i}(\mathbf{L})},
\end{eqnarray} 
where $\lambda_{i}(\mathbf{L})$ indicates the $i$-th eigenvalue of $\mathbf{L}$. The spectral entropy of $\boldsymbol{\rho}$ can be rewritten as 
\begin{eqnarray}
S(\boldsymbol{\rho}) = -\sum_{i=1}^{N}\lambda_{i}(\boldsymbol{\rho})\log_{2}\lambda_{i}(\boldsymbol{\rho}).
\end{eqnarray}
It follows that
\begin{equation}
\label{eq:eigen-relation}
\lambda_{i}(\boldsymbol{\rho})=Z^{-1}e^{-\beta \lambda_{i}(\mathbf{L})} 
\end{equation}
provides the relationship between the eigenvalues of the density and the Laplacian matrices. Using Eq.\,(\ref{eq:eigen-relation}), the spectral entropy of the network $G$ reduces to
\begin{eqnarray}
S(G) &=& \frac{1}{Z\log{2}}\sum_{i=1}^{N}e^{-\beta\lambda_{i}(\mathbf{L})}[\log Z+\beta\lambda_{i}(\mathbf{L})]\nonumber\\
&=&\log_{2} Z-\beta\frac{\partial \log_{2}Z}{\partial \beta}=\log_{2} Z + \beta\Tr{[\mathbf{L}\boldsymbol{\rho}]}.
\end{eqnarray}

A possible interpretation of this entropy will be given later. Here, it is worth discussing some very special cases, where the spectral entropy following from our definition provides very different results from BGS ($S_{bgs}=S(\boldsymbol{\rho}_{bgs})$). 

\vspace{0.25truecm}\noindent\emph{Networks of isolated nodes.} First, let us consider the case of a network with no links among nodes. The eigenvalues of the combinatorial Laplacian are all 0, $Z=N$ and $S=\log_{2}N$, i.e.~the entropy is maximum regardless of $\beta$---whereas $S_{bgs}$ is not defined.

\begin{figure*}[!t]
\centering
\includegraphics[width=\textwidth]{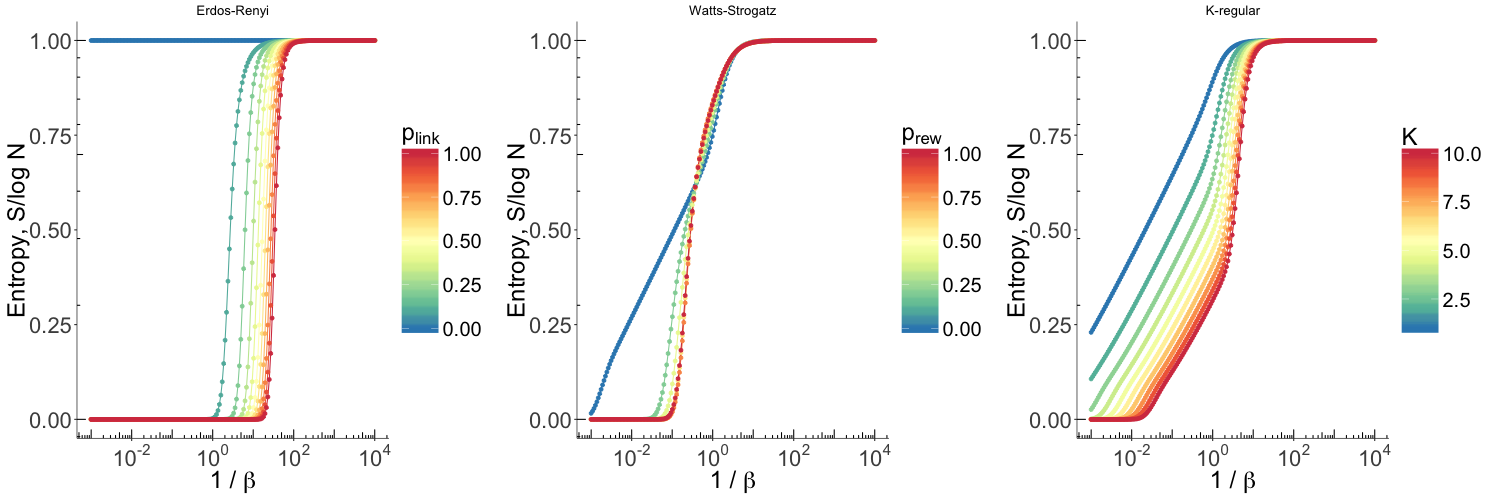}
\caption{\textbf{von Neumann entropy of complex network models.} Spectral entropy as a function of $1/\beta$ for Erd\"os-R\'{e}nyi networks (left-hand panel), Watts-Strogatz's small-world networks (central panel) and K-regular lattices (right-hand panel). Color encodes realizations obtained by varying the parameter of the network model, i.e. the probability $p_{link}$ of a link between two nodes, the rewiring probability $p_{rew}$ and the number of neighbors $K$, respectively. In all cases, networks with $N=200$ are considered.}\label{fig:entropy-toy}
\end{figure*}

\vspace{0.25truecm}\noindent\emph{Single-link networks.} Let us consider a network with only one link between node $i$ to node $j$, i.e. $\mathbf{A}=\mathbf{E}(ij) + \mathbf{E}(ji)$ being $\mathbf{E}(ij)$ the canonical matrix with all zero entries except in $i-$th-$j-$th component. It is straightforward to show that the eigenvalues of the combinatorial Laplacian are all 0 except one whose value is 2. While $S_{bgs}=0$, it follows that
\begin{eqnarray}
S = \log_{2}{Z}+\frac{2\beta e^{-2\beta}}{Z\log{2}}
\end{eqnarray}
where $Z=N-1 + e^{-2\beta}$ and the asymptotic behavior is $\log_{2}{N}$ for any $\beta$. Our spectral entropy scales with the size of the network when only one directed link is present.

\vspace{0.25truecm}\noindent\emph{Complete networks.} In the case of the clique network, there are $N-1$ eigenvalues equal to $N$, from which
\begin{eqnarray}
S =  \log_{2}{Z} + \frac{\beta N (Z-1)}{Z\log{2}}
\end{eqnarray}
where $Z=1+(N-1)e^{-\beta N}$ and the asymptotic behavior is $\log_{2}{N}$ for small $\beta$ and 0 for large $\beta$. 

\vspace{0.25truecm}\noindent\emph{Complex network models.} We show in Fig.\,\ref{fig:entropy-toy} the von Neumann entropy as a function of $1/\beta$ for Erd\"os-R\'{e}nyi~\cite{erdos1959random}, Watts-Strogatz~\cite{watts1998collective} and K-regular networks, for different values of their parameters, the link probability $p_{link}$, the rewiring probability $p_{rew}$ and the number $K$ of node neighbors, respectively. Intriguingly, in all three network types, when $\beta$ is high the entropy tends to zero whereas it approaches the theoretical limit $\log_{2} N$ when $\beta$ is small enough.

\vspace{0.25truecm}\noindent\emph{Sub-additivity of spectral entropy.} Let us consider an undirected network $G$ of $N$ nodes changing over time, with adjacency matrix $\mathbf{A}(0)$ at time $t=0$. At each time step $t$ a pair of two nodes, chosen uniformly randomly and not yet connected, is linked by one undirected link. This is equivalent to having another network $G'(t)$ consisting of just one link and $N-2$ isolated nodes, whose adjacency matrix $\mathbf{A}'(t)$ is summed up to $\mathbf{A}(t-1)$ such that $\mathbf{A}(t) = \mathbf{A}(t-1) + \mathbf{A}'(t)$. If $G(t-1)$ and $G'(t)$ have no edges in common, the above operation is equivalent to the union of the two graphs, another typical approach to aggregate networks.

\begin{figure}[!h]
\centering
\includegraphics[width=.5\textwidth]{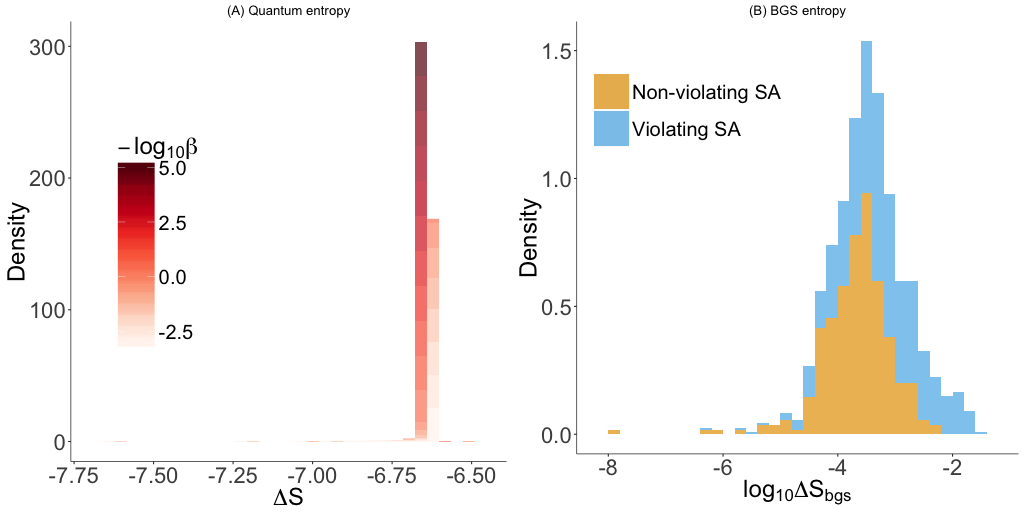}
\caption{\textbf{Sub-additivity of spectral entropy.} Distribution of entropy variation (see the text) for an Erd\"os-R\'{e}nyi network with $p_{link}=0.01$ and $N=100$. The results for the quantum entropy defined in this work (left-hand panel) and the BGS entropy (right-hand panel) are shown. Quantum entropy never violates sub-additivity, regardless of the $\beta$ parameter, whereas the BGS entropy significantly violates sub-additivity.}\label{fig:entropy-changes-ER}
\end{figure}

One immediately asks if the spectral entropy defined in this work and the BGS entropy satisfy the sub-additivity property such that 
\begin{eqnarray}
S(G(t)) \leq S(G(t-1)) + S(G'(t)).
\end{eqnarray}
We show in Fig.\,\ref{fig:entropy-changes-ER} the distribution of $\Delta S = S(G(t)) - [S(G(t-1)) + S(G'(t))]$ obtained by calculating both quantum and BGS entropy for an ensemble of Erd\"os-R\'{e}nyi networks. Similar results are obtained by using ensembles of Watts-Strogatz and K-regular networks.

As proven in the Appendix, the spectral entropy does not violate the sub-additivity property regardless of the value of $\beta$, whereas the $S_{bgs}$ often violates these sensible additivity requirement. See Appendix~\ref{app:subadd} for further details.

\section{R\'{e}nyi entropy of complex networks}

The von Neumann entropy is just a special case of the more general R\'{e}nyi entropy of a quantum system, defined by~\cite{wehrl1978general}
\begin{eqnarray}
S_{q}=\frac{1}{1-q}\log_{2} \Tr \boldsymbol{\rho}^{q} = \frac{1}{1-q}\log_{2} \sum_{i=1}^{N}\lambda_{i}(\boldsymbol{\rho})^{q}.
\end{eqnarray}
It is widely used in quantum information theory and quantum computing to quantify entanglement~\cite{2010RvMP...82..277E} and correlations in physical systems~\cite{2008JPhA...41b5302F}, it can be expressed as families of tensor network contractions~\cite{2013JPhA...46U5301B} and has been found to share a close relationship to free energy~\cite{2011arXiv1102.2098B}.

By using Eq.~(\ref{eq:eigen-relation}), the R\'{e}nyi spectral entropy of a complex network is therefore given by
\begin{eqnarray}
S_{q}=\frac{1}{1-q}\ln \sum_{i=1}^{N}Z^{-q}e^{- q\beta \lambda_{i}(\mathbf{L})},
\end{eqnarray}
in terms of the eigenvalues of the Laplacian matrix. In fact, this entropy generalizes other entropic measures. As $q$ approaches $0$, $S_{q}$ increasingly weights all eigenvalues more equally, approaching the Hartley entropy. In the limit of $q\rightarrow1$ R\'{e}nyi entropy approaches the spectral entropy. As $q$ approaches $\infty$, $S_{q}$ becomes dominated by the high-probability events and it converges to the min-Entropy. The case with $q=2$, recovers the collision entropy---in the case of quantum systems, this is connected to the purity of the system. In fact, $S_{2}$ identically vanishes for $\boldsymbol{\rho}^2=\boldsymbol{\rho}$, i.e.~for pure states.

\begin{figure*}[!t]
\centering
\includegraphics[width=.99\textwidth]{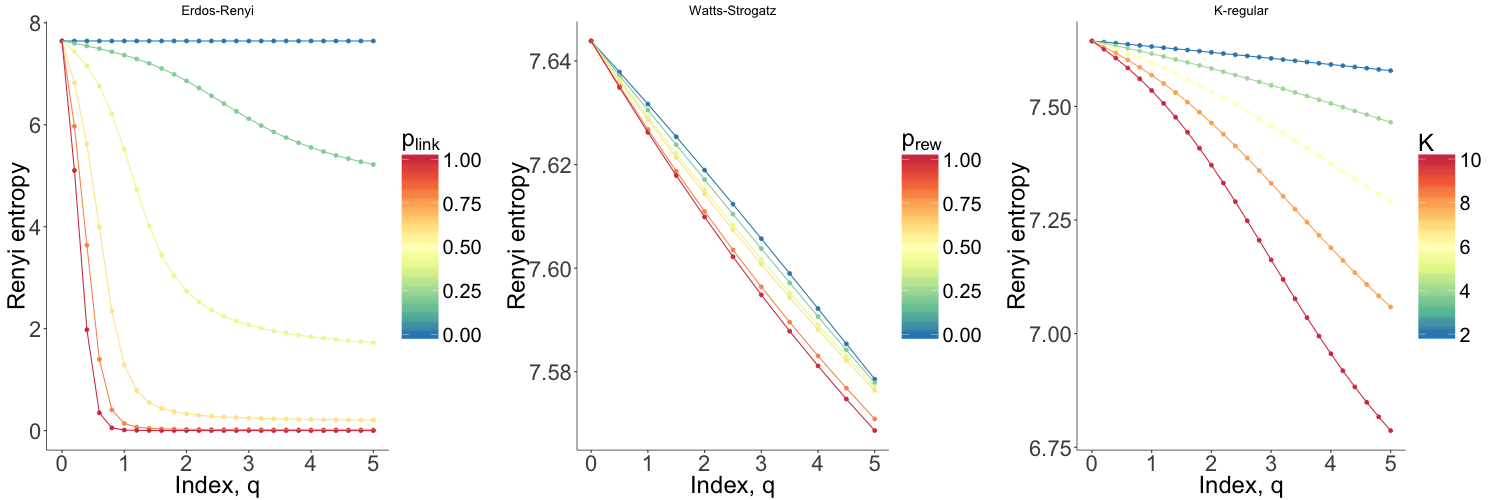}
\caption{\textbf{R\'{e}nyi entropy of complex network models.} Entropy as a function of $q$ with $\beta=1/15.8$ for Erd\"os-R\'{e}nyi networks (left-hand panel), Watts-Strogatz's small-world networks (central panel) and K-regular lattices (right-hand panel). Color encodes realizations obtained by varying the parameter of the network model, i.e. the probability $p_{link}$ of a link between two nodes, the rewiring probability $p_{rew}$ and the number of neighbors $K$, respectively. In all cases, networks with $N=200$ are considered.}\label{fig:entropy-renyi-er}
\end{figure*}

We show in Fig.\,\ref{fig:entropy-renyi-er} entropy as a function of $q$ at different values of $\beta$ for Erd\"os-R\'{e}nyi, Watts-Strogatz and K-regular networks, respectively, for different values of their parameters. It is common to assess that the Watts-Strogatz network reduces to a K-regular graph for $p_{rew}=0$ and approaches an Erd\"os-R\'{e}nyi network for $p_{res}=1$. However, the behavior of R\'{e}nyi entropy as a function of $q$ shows that there are some significant differences at least in the latter case, where the entropy is considerably larger for the Watts-Strogatz model than the Erd\"os-R\'{e}nyi one.

\section{Generalized quantum divergences between two complex networks}

One of the main goal of information theory is to quantify the amount of information about a probability distribution -- generally obtained from empirical measurements -- provided that one has full information about another probability distribution -- e.g. the model. This goal is achieved by introducing relative entropies or, equivalently, information divergences.

Similarly, the introduction of divergences (a.k.a.~quantum relative entropy) in quantum information theory is foundational to the quest to understand differences between quantum states, quantum and classical information, the quantification of the thermodynamic cost of communication as well as optimal protocols to transfer information---see e.g.~the reviews \cite{vedral2002role} and \cite{schumacher2002relative}.

The quantum R\'{e}nyi entropy can be used to define the quantum R\'{e}nyi divergence, also known as $q-$relative R\'{e}nyi entropy, by
\begin{eqnarray}
\mathcal{D}_{q}(\boldsymbol{\rho}||\boldsymbol{\sigma})=\frac{1}{q-1}\log_{2}\Tr{(\boldsymbol{\rho}^{q}\boldsymbol{\sigma}^{1-q})}
\end{eqnarray}
which reduces to the quantum Kullback-Leibler divergence (or, equivalently, the quantum relative entropy)
\begin{eqnarray}
\label{eq:kld-quantum}
\mathcal{D}_{1}(\boldsymbol{\rho}||\boldsymbol{\sigma})=\Tr{[\boldsymbol{\rho}(\log_{2}\boldsymbol{\rho}-\log_{2}\boldsymbol{\sigma})]}
\end{eqnarray}
for $q\rightarrow1$. 

In general such divergences are not symmetric and bounded, making difficult certain comparison uses. An alternative measure is the quantum $q-$Jensen-Shannon divergence~\cite{majtey2005jensen}
\begin{eqnarray}
\mathcal{J}_{q}(\boldsymbol{\rho}||\boldsymbol{\sigma})= S_{q}\left(\frac{\boldsymbol{\rho}+\boldsymbol{\sigma}}{2}\right) - \frac{1}{2}[S_{q}(\boldsymbol{\rho})+S_{q}(\boldsymbol{\sigma})],
\end{eqnarray}
where $\boldsymbol{\mu}=\frac{\boldsymbol{\rho}+\boldsymbol{\sigma}}{2}$ is usually called mixture matrix. It can be proven that $\mathcal{J}_{q}^{\frac{1}{2}}$ defines a true metric for $0\leq q<2$~\cite{majtey2005jensen,lamberti2008metric,briet2009properties} and can be used as a measure of distinguishability~\cite{dedomenico2015structural} or similarity~\cite{gerlach2016similarity}.

Our formalism allows to use the family of $q-$quantum divergences to compare two networks and this opens up a new approach to attack two fundamental problems in network science.

\subsection{Maximum Likelihood Estimation and Model Selection}

For a given model and its corresponding set of parameters, the likelihood function measures the probability of observing the data according to the model parameters. Therefore, it is sufficient to perform maximum likelihood estimation to obtain the parameters of the model that better reproduce the data, according to the model.

While it is straightforward to define the likelihood function for probability distributions, it is challenging to define a similar concept in the context of density matrices. In the case of complex networks the challenge is complicated by the lack of an appropriate probability distribution. In fact, one usually designs a model, or a set of models, to reproduce only some salient features of an observed complex network, often limiting the comparison between model and data to a few descriptors such as degree distribution, degree correlations, clustering or path statistics. 

Quantum divergences provide a grounded and unifying approach to this problem. Let us start from the classical problem where an empirical probability distribution $P(x)$ is obtained from observing the $N$ outcomes $\{x_{i}\}$ ($i=1,2,...,N$) of a stochastic variable $\mathcal{X}$. Let $Q(x;\Theta)$ be a model to approximate $P(x)$, which depends on one (or more) parameter(s), here indicated by $\Theta$. In this context, the Kullback-Leibler divergence measures the information gain when the model $Q(x;\Theta)$ is used to explain the observation $P(x)$, and it can be written as
\begin{eqnarray}
\label{eq:kld-class}
\mathcal{D}(P||Q)&=&\int~dx P(x)\log_{2}\frac{P(x)}{Q(x;\Theta)}\nonumber\\
&=& -S(P) - \int~dx P(x)\log_{2}Q(x;\Theta).
\end{eqnarray}

If the model $Q$ is plausible, there exist a value $\Theta^{\star}$ such that the divergence is minimum and we are interested in finding such a value by minimizing the divergence with respect to $\Theta$. We notice that the first term in the right-hand side of Eq.~(\ref{eq:kld-class}) does not depend on $\Theta$ and therefore plays no role in the minimization procedure. By noticing that 
\begin{eqnarray}
P(x)=\frac{1}{N}\sum_{i=1}^{N}\delta(x-x_{i}),\nonumber
\end{eqnarray}
where $\delta$ is the Dirac function, the second term in the right-hand side of Eq.~(\ref{eq:kld-class}) reduces to
\begin{eqnarray}
\int dx~P(x)\log_{2}Q(x;\Theta)=\frac{1}{N}\sum_{i=1}^{N}\log_{2}Q(x_{i}; \Theta),
\end{eqnarray}
which is proportional to the negative log-likelihood function. Here, the pre-factor can be safely neglected during the minimization procedure. Therefore, by minimizing the Kullback-Leibler divergence one effectively maximizes the log-likelihood function:
\begin{eqnarray}
\label{eq:lik-class}
\min\limits_{\Theta}\{\mathcal{D}(P||Q)\} = \max\limits_{\Theta}\{\log_{2}\mathcal{L}(x;\Theta)\}.
\end{eqnarray}

We use the proposed framework to achieve a similar result in the case of density matrices. Let $\boldsymbol{\rho}$ be the density matrix of an empirical network and let $\boldsymbol{\sigma}(\Theta)$ a model for such a network, depending on one (or more) parameter(s), here indicated by $\Theta$. By starting from Eq.~(\ref{eq:kld-quantum}), and by arguments similar to the classical case, it is straightforward to show that
\begin{eqnarray}
\label{eq:lik-quantum}
\min\limits_{\Theta}\{\mathcal{D}(\boldsymbol{\rho}||\boldsymbol{\sigma})\} = \max\limits_{\Theta}\left\{\Tr{[\boldsymbol{\rho}\log_{2}\boldsymbol{\sigma}(\Theta)]}\right\}.
\end{eqnarray}
By comparing the right-hand side of Eq.~(\ref{eq:lik-class}) and Eq.~(\ref{eq:lik-quantum}), we define the \emph{network log-likelihood function} by
\begin{eqnarray}
\label{eq:ll}
\log_{2}\mathcal{L}(\Theta)=\Tr{[\boldsymbol{\rho}\log_{2}\boldsymbol{\sigma}(\Theta)]},
\end{eqnarray}
where the likelihood function can be calculated by exploiting the properties of the matrix exponential as
\begin{eqnarray}
\label{eq:likelihood}
\mathcal{L}(\Theta)&=&e^{\Tr{[\boldsymbol{\rho}\log\boldsymbol{\sigma}(\Theta)]}}\nonumber\\
&=&\det\left(e^{\boldsymbol{\rho}\log\boldsymbol{\sigma}(\Theta)}\right).
\end{eqnarray}

This result allows us to obtain a maximum-likelihood estimation of parameter(s) $\Theta=\{\theta_{1}, \theta_{2}, ..., \theta_{d}\}$ by minimizing the Kullback-Leibler divergence between networks. The covariance matrix corresponding to this estimation is the Fisher information matrix (see Appendix~\ref{app:fisher}), whose classical counterpart is equivalent to the Hessian of the Kullback-Leibler divergence and it is used to assess the quality of the spectral likelihood estimate. If $\Theta^{\star}$ is an unbiased estimator of $\Theta$, the associated covariance matrix satisfies the Cramer-Rao bound
\begin{eqnarray}
\text{cov}(\Theta^{\star})\geq \boldsymbol{\mathcal{I}}^{-1}(\Theta^{\star}).
\end{eqnarray}

We can exploit this finding to go beyond model fitting, defining an operative procedure for model selection.

In fact, generally more than one model is used to understand the data and a fundamental problem in data analysis, known as model selection, is to quantify which model out of a set of candidates is the best one in reproducing the data. One solution to this problem has been given by Akaike, who proposed an information criterion (AIC) for this purpose~\cite{akaike1973}, by showing that the expected value of the relative cross-entropy term in the Kullback-Leibler divergence equals the log-likelihood of the model given the data plus a penalizing constant term which accounts for the number of free parameters. The AIC is given by
\begin{eqnarray}
\label{eq:AIC}
\text{AIC} = 2k - 2\log_{2}\mathcal{L}(\Theta^{\star})
\end{eqnarray}
where $k$ is the number of parameters of the model and we plug Eq.~(\ref{eq:ll}) into Eq.~(\ref{eq:AIC}) for applications to complex networks. In practice, given a set of models $\mathcal{M}=\{M_{1}, M_{2}, ..., M_{n}\}$, with number of parameters $k_{1}, k_{2}, ..., k_{n}$ and likelihood $\mathcal{L}_{1}, \mathcal{L}_{2}, ..., \mathcal{L}_{n}$, respectively, the most suitable candidate to explain the data is the one being a trade-off between having as small as possible divergence from the data and as small as possible number of parameters, i.e.~the one such that $\text{AIC}$ is minimum.

Similarly, other model selection criteria can be extended from information theory to the complex network framework. This is the case of Bayesian information criterion (BIC) defined by
\begin{eqnarray}
\label{eq:BIC}
\text{BIC} = k\log_{2}N - 2\log_{2}\mathcal{L}(\Theta^{\star})
\end{eqnarray}
and Fisher information approximation (FIA) defined by 
\begin{eqnarray}
\label{eq:FIA}
\text{FIA} &=& \frac{k}{2}\log_{2}\frac{N}{2\pi} - \log_{2}\mathcal{L}(\Theta^{\star})\nonumber\\
&+&\log_{2}\left(\int~d\Theta \sqrt{\det{\boldsymbol{\mathcal{I}}(\Theta)}}\right),
\end{eqnarray}
where the last term penalizes a model because of its geometric complexity~\cite{rissanen1996fisher,pitt2002toward}, a typical concept in information geometry~\cite{amari2007methods}. In classical statistical inference, it has been shown~\cite{rissanen1996fisher} that the FIA quantifies the length of the shorted description of the data given the model and, as a consequence, its minimization corresponds to finding the minimum description length (MDL) of the network. The MDL principle~\cite{rissanen1978modeling}, also known as a formalization of Occam's razor, represents data and models as codes to be compressed, where the model better compresses data to provide its best description. This principle is one of the most important concepts in information theory and its interpretation, in our context, opens the door for future studies.

\begin{figure*}[!t]
\centering
\includegraphics[width=.95\textwidth]{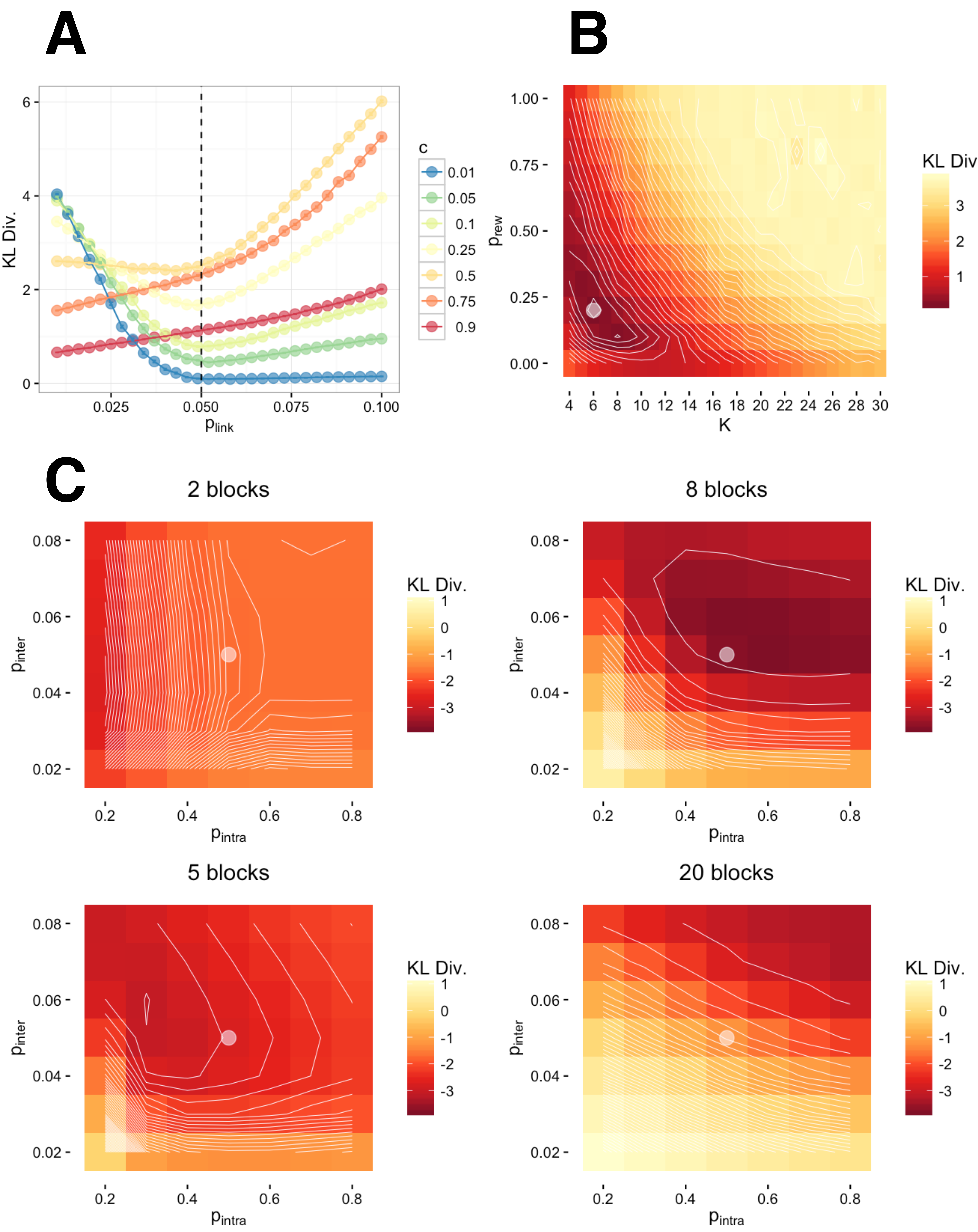}
\caption{\textbf{Maximum-likelihood parameter estimation based on Kullback-Leibler minimization.} (A) Erd\"os-R\'{e}nyi network model ($N=200$) for different values of the link probability $p_{link}$ against an Erd\"os-R\'{e}nyi network with $p_{link}^{\star}=0.05$, assumed to be the data to fit. The vertical dashed line indicates the true value of the parameter. Each curve corresponds to a value $\beta^{\star}$ of $\beta$ such that $S[\boldsymbol{\rho}(\beta^{\star})]/\log_{2}N=c(\beta^{\star})$, with $c$ a real number between 0 and 1. Global minima are expected around the true value. (B) Watts-Strogatz's network model ($N=200$) for different values of the parameters $K$ and $p_{rew}$ against a Watts-Strogatz network with $K^{\star}=6$ and $p_{rew}^{\star}=0.2$, assumed to be the data to fit. The white dot indicates the true value and it falls in the iso-likelihood region with smallest Kullback-Leibler divergence (encoded by color in log scale). (C) Stochastic block-model ($N=200$) for different values of the intra- and inter-community probability parameters, $p_{intra}$ and $p_{inter}$, and number of equally sized blocks against a network obtained from the same model with 8 blocks, $p_{intra}^{\star}=0.5$ and $p^{\star}_{inter}=0.05$, assumed to be the data to fit.}\label{fig:model-selection}
\end{figure*}

To probe the power of the proposed method we performed several tests on synthetic networks. We generate a network from a specific model with a given value of the parameter(s), assuming it is the observed network, and a set of networks from the same model by varying the value of the parameter(s). Therefore, we perform a maximum-likelihood parameter estimation based on Kullback-Leibler minimization to find the value of the parameter(s) that better fit the observation.

In Fig.\,\ref{fig:model-selection}A we consider the case of an Erd\"os-R\'{e}nyi network model ($N=200$), being link probability $p_{link}$ the parameter to fit and $p_{link}^{\star}=0.05$ its expected value. By scanning over the range allowed for $p_{link}$, we sample an ensemble of 100 realizations for each value of the link probability and compare each network against the observed one to calculate the average Kullback-Leibler divergence. In principle, the procedure depends on the value of $\beta$, therefore we perform the analysis for different values of this parameter in order to understand which value (or range of values) provides the best fit. We consider specific values of $\beta$ for this purpose, more specifically we calculate the $\beta^{\star}$ such that the entropy normalized to its maximum value, i.e.~$\log_{2}N$, gets a specific real value $c(\beta^{\star})$ between 0 and 1. We choose values for $c(\beta^{\star})$ ranging between 0.01 and 0.9, and the results show that the global minima corresponds or are very close to the expected value, for $c(\beta^{\star})\leq 0.5$. The best performance is obtained for $c(\beta^{\star})< 0.1$. This analysis suggests a rule of thumb to choose a specific value of $\beta$ for fitting purposes: the region close to the critical point -- where entropy changes from 0 to a positive value -- provides the most performant range for $\beta$. 

In Fig.\,\ref{fig:model-selection}B we show the Kullback-Leibler divergence of Watts-Strogatz's network model ($N=200$) for different values of the parameters $K$ and $p_{rew}$ against a Watts-Strogatz network with $K^{\star}=6$ and $p^{\star}_{rew}=0.2$, assumed to be the empirical data. The result shows that the most likely region of the parameter space, i.e. the one where the model is more informative about the data, is successfully identified by the Kullback-Leibler minimization procedure previously described. The result is very interesting because only a single realization of the model, instead of an ensemble as in the previous case, has been used for each pair of parameters, suggesting that the procedure is robust against sample size while reducing the computational cost of the calculation.

Fig.\,\ref{fig:model-selection}C shows the Kullback-Leibler divergence of a stochastic block-model~\cite{holland1983stochastic} ($N=200$) for different values of the intra- and inter-community probability parameters, $p_{intra}$ and $p_{inter}$, and number of equally sized blocks against a network obtained from the same model with 8 blocks, $p_{intra}^{\star}=0.5$ and $p^{\star}_{inter}=0.05$, assumed to be the empirical data. As for the Erd\"os-R\'{e}nyi case, we sample an ensemble of 100 realizations for each triad of parameters, to calculate the average Kullback-Leibler divergence between model and a observation. We show the results for different block sizes and the overall minimum is found for 8 blocks, $p_{intra}=0.6$ and $p_{inter}=0.05$, in excellent agreement with expectation.

\begin{figure*}[!t]
\centering
\includegraphics[width=.95\textwidth]{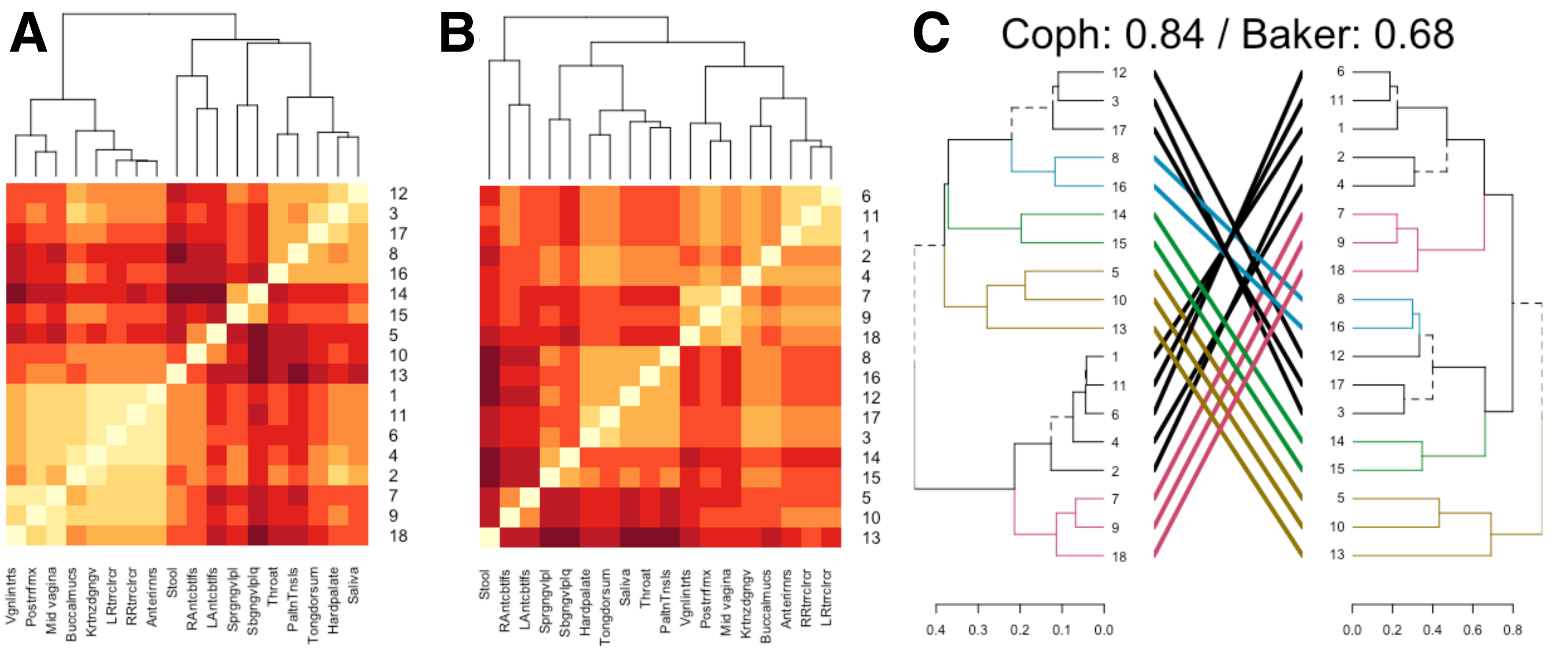}
\caption{\textbf{Hierarchical clustering of human microbiome sites.} The Jensen-Shannon distance matrices with (A) $\beta=0.1$ and (B) $\beta=10$ are shown, together with their (C) tantlegram, i.e., a visual analysis of their differences. Cophenetic distance and Baker's gamma correlation coefficient, quantifying the correlation between the two dendrograms, are reported.}\label{fig:cluster-multiplex}
\end{figure*}

\subsection{Clustering layers of multilayer systems}

The second application concerns the more recent problem of identifying layers of a multiplex network or snapshots of a time-varying network~\cite{dedomenico2013mathematical} that provide redundant information about the system and that might be aggregated to reduce the complexity of the system~\cite{dedomenico2015structural}. In this case, by using appropriate quantum divergences it is possible to calculate a distance between layers or snapshots and devise exact or heuristics procedures to cluster them. With the recent rise of interest in multilayer systems~\cite{kivela2013multilayer,boccaletti2014structure,dedomenico2016physics}, this problem became very relevant for practical applications, e.g. to demonstrate the validity of a multilayer approach to modeling complex systems as the human brain~\cite{dedomenico2016mapping}, and new approaches to cluster and aggregate layers have been proposed more recently~\cite{stanley2016clustering,taylor2016enhanced}.

Here, we use the networks built from analysis of the structure and function of the human microbiome, consisting of 18 layers of a multiplex network, each one corresponding to a body site. Recently such layers have been partitioned into community types, by using Dirichlet multinomial mixture models, that may be associated with complex diseases~\cite{ding2014dynamics}. We use the same data and calculate the Jensen-Shannon distance between each pair of layers for different values of the $\beta$ parameter. We show in Fig.\,\ref{fig:cluster-multiplex} the resulting hierarchical clustering of the layers, in good agreement with the result reported in~\cite{ding2014dynamics}. A more quantitative comparison against the results in~\cite{ding2014dynamics} is difficult because only $p-$value upper bounds are reported in the study for community-type associations. Nevertheless, our method is able to correctly reproduce the clustering of sites within certain body regions. For instance, the community consisting of hard palate, tongue dorsum, saliva, palatine tonsils and throat, as well as the community made by vaginal introitus, mid vagina and posterior fornix and the smaller ones made by sub- and supra-gingival plaque and left/right retroauricular crease. For ranges of $\beta$ within one order of magnitude, the differences between hierarchies are small and not significant. Here we show the results for $\beta=0.1$ and $\beta=10$, with cophenetic distance equal to 0.84 and Baker's gamma correlation coefficient equal to 0.68. By performing the same analysis on hierarchies where labels are reshuffled uniformly random, we reject the null hypothesis that the two dendrograms are uncorrelated ($P<0.001$), supporting the fact that the result is robust to the choice of $\beta$.

\section{Discussion}

The use of von Neumann entropy is central to modern quantum information theory, with many new insights, uses and interpretations arising often.  In the context of complex networks, spectral entropy might be considered as a measure of the information content of the system, although this interpretation is not easy to accept without the appropriate formalization typical of classical and quantum information theory.

Here, we observe that networks of isolated nodes (which challenge the notion of a network itself) and fully-connected networks, i.e.~cliques, have maximal entropy. Of course, connectivity is not the feature common to such extremal network scenarios. We investigated several alternative explanations, but the most promising one is based on the number of possible configurations obtained by reshuffling the connections among nodes, that is equivalent to randomly reshuffling the entries of the adjacency matrix. In fact, all possible reshuffled configuration of both the network of isolated nodes and the clique network do not alter the system. In other words, the graph isomorphism problem becomes trivial, as all possible networks are not distinguishable from each other, thus maximizing the uncertainty about the system and, consequently, the entropy.
In the case of K-regular networks the situation is a bit different, because not all reshuffled configurations will keep them unaltered: therefore such networks are expected to exhibit lower entropy than $\log_{2}N$. It is interesting the case of an Erd\"os-R\'{e}nyi network, which is characterized by stochastically homogeneous connectivity. The entropy for this type of networks is almost always zero or $\log_{2}N$, except for a narrow range of $\beta$. This is compatible with our interpretation, because on average almost any reshuffled configuration of this type of networks will resemble the original. 
According to this interpretation, extremely sparse networks and extremely dense networks should exhibit almost maximum entropy. We have performed additional numerical experiments and theoretical analysis of toy models (path graphs, lattices, etc.) and we have verified our expectation.

The proposed framework is entirely based on the computation of eigenvalues of density matrices and their products. The computational complexity of such operation is generally expected to be polynomial and to scale as $\mathcal{O}(n^{\gamma})$, with $2<\gamma<2.4$, making challenging calculations for networks with hundreds of thousands nodes. However, the recent advances in parallel operations on dense symmetric matrices make possible to solve eigenvalue problems for matrices with size of the order of $10^{5}$ in less than five minutes on standard computer machines~\cite{bientinesi2005parallel}.

\section{Conclusions and outlook}

The concept of thermodynamic entropy has been crucial to understand the structure and the dynamics of complex systems. The generalization of this concept to quantum mechanics by von Neumann was a milestone in the field, allowing, among other applications, to characterize the mixedness of quantum states.

An appropriate definition of entropy has been lacking in the field of complex networks, mainly because there is no simple way to define a probability distribution able to represent, without loss of information, the network. Previous attempts to define the entropy of a complex network where mainly based on the calculation of Shannon entropy of the probability distribution of some descriptor(s), while neglecting other information. 

More recently, the idea that a quantum-like entropy might be introduced for complex networks has been explored. Here, we have demonstrated that previous attempts fail in preserving fundamental properties of entropy, like sub-additivity. Motivated by this finding, we have proposed a density matrix, inspired by the Gibbs state of a quantum system, which non-trivially depends on the combinatorial Laplacian matrix of the network. One of the main advantages of this approach, rooted in spectral theory, is that the resulting entropy does not depend on the distribution of a specific network descriptor but it depends on the network as a whole.

In this study, we have shown, analytically and numerically, that our density matrix allows to preserve the basic properties of an entropy. More specifically, we have demonstrated that the entropy of the system obtained by aggregating two different networks must be equal to or smaller than the sum of the entropy of the two non-aggregated networks. 

This definition of entropy has allowed us to define R\'enyi spectral entropy and generalized relative entropies, also known as quantum divergences. By using Kullback-Leibler divergence and well-known results in classical information theory we have devised a maximum-likelihood approach to model fitting, entirely based on the eigenvalue spectra of density matrices. Our numerical experiments confirmed that by minimizing the Kullback-Leibler divergence between observed network and models, we obtain the maximum spectral likelihood estimation of parameters. This result might be of special interest for applications devoted to unravel the meso-scale organization of a network, among others.

Finally, by using the Jensen-Shannon divergence, whose square root provides a metric, we have shown that our framework can be applied to quantify the spectral distance between two networks. More specifically, this result is of interest for applications in multilayer network research, where the distance among layers can be measured and used to hierarchically cluster them. As a representative example of the power of our methodology, we have shown that this approach recovers with excellent accuracy the community-based classification of human microbiome sites.

This study opens interesting perspectives on future works. For instance, while our inference procedure does not provide the community label to assign to each node, as in community-detection methods based on stochastic block-modeling~\cite{guimera2009missing,karrer2011stochastic,decelle2011asymptotic,peixoto2012evolution,peixoto2013parsimonious,peixoto2015model,newman2016structure}, it provides a fast way, based on network's spectral properties, to explore the parameter space in order to identify the most informative region about the data. An intriguing possibility to explore is how this result can complement existing network inference approaches~\cite{zdeborova2015statistical}.

A future application of our framework---of great interest for the community of network scientists---is the quantification of how much information is needed for correctly learning the parameters of a model, that should not be confused with the Cramer-Rao bound. This represents a crucial issue in network science: connectivity of empirical networks is often sampled through \emph{ad hoc} algorithms, as in the case of virtual social systems, the Internet or the World Wide Web. Quantifying to which extent it is possible to learn the parameter(s) of a network model in absence of partial connectivity information is a tantalizing possibility, expected to deepen our understanding of networked systems.

The information-theoretic quantities introduced in this work open the intriguing possibility to extend information geometry to complex networks. In information geometry parametric statistical models define a Riemannian manifold endowed with Fisher information matrix as a metric tensor~\cite{rao1945information,jeffreys1946invariant}. The importance of information manifolds for statistical inference is well established and found applications to machine learning~\cite{amari1995information,amari2001information,amari2007methods} and phase transitions in quantum systems~\cite{braunstein1994statistical,zanardi2007information,rezakhani2010intrinsic,rams2011quantum,strobel2014fisher,hauke2016measuring}. It will be interesting to explore to which extent the success of information geometry can be ported to statistical inference of complex networks, where the metric tensor of the underlying information manifold is given by the spectral Fisher information matrix.

The potential of the proposed methodologies goes beyond network science and we envision important contributions to quantum physics, especially to the emergent field of Hamiltonian learning~\cite{granade2012robust,wiebe2014hamiltonian} and inference of quantum complex network models based on qubit entanglement~\cite{cuquet2009entanglement,perseguers2010quantum}.

\begin{acknowledgments}
The authors thanks Alex Arenas for useful discussions. MDD acknowledges financial support from the Spanish program Juan de la Cierva (IJCI-2014-20225). JDB acknowledges IARPA and the Foundational Questions Institute (FQXi) for financial support.
\end{acknowledgments}


\appendix

\section{Sub-additivity of the von Neumann entropy for aggregate networks}\label{app:subadd}


Let $\mathbf{A}$ and $\mathbf{B}$ be the adjacency matrices of two networks $G_{A}$ and $G_{B}$, respectively, with $N$ nodes. Let $\mathbf{C}=\mathbf{A}+\mathbf{B}$ be the adjacency matrix obtained by their sum, which represents a new network $G_{C}$ obtained by aggregating $G_{A}$ and $G_{B}$. In the following we will show that $S(G_{C})\leq S(G_{A})+ S(G_{B})$, i.e.~that the sub-additivity property of the entropy is always satisfied. For sake of simplicity, we will use the notation $S_{X}=S(G_{X})$.

Let us indicate with $\mathbf{L}_{A}$, $\mathbf{L}_{B}$ and $\mathbf{L}_{C}$ the Laplacian matrices of $G_{A}$, $G_{B}$ and $G_{C}$, respectively, and let $\boldsymbol{\rho}_{A}$, $\boldsymbol{\rho}_{B}$ and $\boldsymbol{\rho}_{C}$ indicate the corresponding density matrices defined as in Eq.~(\ref{eq:def-rho}).

The Kullback-Leibler divergence of two density matrices is always non-negative, as a consequence of Klein'€™s inequality with $f(\mathbf{X})=\mathbf{X}\ln \mathbf{X}$. Therefore
\begin{eqnarray}
\mathcal{D}(\boldsymbol{\rho}_{C} ||\boldsymbol{\rho}_{A}) &=& - S_{C} - \Tr{[\boldsymbol{\rho}_{C}\log_{2} \boldsymbol{\rho}_{A}] }\nonumber\\
&=& -S_{C} + \beta\Tr{[\mathbf{L}_{A}\boldsymbol{\rho}_{C}]} + \log_{2}Z_{A} \geq 0,
\end{eqnarray}
and similarly for $\mathcal{D}(\boldsymbol{\rho}_{C} ||\boldsymbol{\rho}_{B})$. As $\mathbf{L}_A$, $\mathbf{L}_B$, $\boldsymbol{\rho}_{A}$ and $\boldsymbol{\rho}_{B}$ are positive semi-definite, from their Cholesky factorization it is straightforward to show that
\begin{eqnarray}
\Tr{[\mathbf{L}_{X}\boldsymbol{\rho}_{X}]} &=& \Tr{[(\mathbf{D}\mathbf{D}^{\dagger}) (\mathbf{Q}\mathbf{Q}^{\dagger}])} \nonumber\\
&=&\Tr{[(Q^{\dagger}D)(Q^{\dagger}D)^{\dagger}]}\geq 0.
\end{eqnarray}

Moreover, it is possible to show that $\log_{2} Z_X \geq 0$. In fact, $Z_X=\sum\limits_{i=1}^{N}e^{-\beta\lambda_{i}(\mathbf{L}_{X})}$, and $\lambda_{1}(\mathbf{L}_{X})=0$ because of the Perron-Frobenius theorem. Therefore
\begin{eqnarray}
Z_X=1+\sum\limits_{i=2}^{N}e^{-\beta\lambda_{i}(\mathbf{L}_{X})}\geq 1, 
\end{eqnarray}
from which $\log_{2} Z_X \geq 0$.

By summing up such non-negative terms, the following inequality
\begin{eqnarray}
\mathcal{D}(\boldsymbol{\rho}_{C} ||\boldsymbol{\rho}_{A}) + \mathcal{D}(\boldsymbol{\rho}_{C} ||\boldsymbol{\rho}_{B}) + \nonumber\\
+ \beta\Tr{[\mathbf{L}_{A}\boldsymbol{\rho}_{A}]} + \beta\Tr{[\mathbf{L}_{B}\boldsymbol{\rho}_{B}]} + \log_{2} Z_C \geq 0
\end{eqnarray}
holds. The above inequality can be expanded into 
\begin{eqnarray}
-S_C + \beta\Tr{[\mathbf{L}_{A}\boldsymbol{\rho}_{C}]} +  \log_{2}Z_{A} + \nonumber\\
-S_C + \beta\Tr{[\mathbf{L}_{B}\boldsymbol{\rho}_{C}]} +  \log_{2}Z_{B} + \nonumber\\
\beta \Tr{[\mathbf{L}_{A}\boldsymbol{\rho}_{A}]} + \beta \Tr{[\mathbf{L}_{B}\boldsymbol{\rho}_{B}]} + \log_{2} Z_C\geq0.
\end{eqnarray}
By exploiting the fact that, for a Gibbs-like state, the von Neumann entropy is given by
\begin{equation}
S(\boldsymbol{\rho}_{X})=  \beta \Tr{[\mathbf{L}_{X}\boldsymbol{\rho}_{X}]} +  \log_{2}Z_{X}
\end{equation}
and $\mathbf{L}_C=\mathbf{L}_A+\mathbf{L}_B$, it follows that
\begin{eqnarray}
S_{A} + S_{B} -2S_{C} + \log_{2} Z_C + \beta\Tr{[\mathbf{L}_C\boldsymbol{\rho}_C]}\geq 0,\nonumber
\end{eqnarray}
which leads to $S_{A} + S_{B} \geq S_{C}$.


\section{Spectral Fisher information matrix}\label{app:fisher}

Let $\boldsymbol{\rho}(\Theta)$ the parametric statistical model, depending on the parameter set $\Theta=\{\theta_{1}, \theta_{2}, ...,\theta_{d}\}$, for a complex network with density matrix $\boldsymbol{\rho}$. In quantum physics, the upper bound to the expected Fisher information is given by the quantum information~\cite{helstrom1967minimum,braunstein1994statistical,monras2010information}
\begin{eqnarray}
\mathcal{I}_{\alpha\beta}(\Theta)=\mathbb{E}\left( \boldsymbol{\Sigma}^{(\alpha)}\circ\boldsymbol{\Sigma}^{(\beta)} \right)=\Tr{\left[\boldsymbol{\rho}~\boldsymbol{\Sigma}^{(\alpha)}\circ\boldsymbol{\Sigma}^{(\beta)}\right]},
\end{eqnarray}
where $\boldsymbol{\Sigma}^{(\alpha)}$ is the symmetric logarithmic derivative---with respect to the $\alpha-$th parameter ($\alpha=1,2,...,d$)---defined by 
\begin{eqnarray}
\label{eq:SLD}
\frac{\partial}{\partial\theta_{\alpha}}\boldsymbol{\rho}(\Theta)=\boldsymbol{\Sigma}^{(\alpha)}\circ\boldsymbol{\rho}(\Theta),
\end{eqnarray}
and we have used the symmetric product defined by
\begin{eqnarray}
\mathbf{X}\circ\mathbf{Y}=\frac{1}{2}(\mathbf{X}\mathbf{Y}+\mathbf{Y}\mathbf{X}).
\end{eqnarray}
Let us consider the spectral decomposition of the density matrix
\begin{eqnarray}
\boldsymbol{\rho}(\Theta)=Z(\Theta)^{-1}\mathbf{Q}(\Theta)e^{-\beta\boldsymbol{\Lambda}(\Theta)}\mathbf{Q}^{-1}(\Theta), 
\end{eqnarray}
where $\boldsymbol{\Lambda}(\Theta)$ is the diagonal matrix of Laplacian's eigenvalues $\lambda_{i}(\Theta)$ ($i=1,2,...,N$) and the columns of the matrix $\mathbf{Q}(\Theta)$ are the corresponding eigenvectors, that we indicate by $\mathbf{q}_{i}(\Theta)$. Note that we do not indicate explicitly the dependence on $\mathbf{L}$ for sake of simplicity. In the Laplacian eigenbasis, Eq.~(\ref{eq:SLD}) reads
\begin{eqnarray}
\mathbf{q}^{\top}_{i}(\Theta)\frac{\partial}{\partial\theta_{\alpha}}\boldsymbol{\rho}(\Theta)\mathbf{q}_{j}(\Theta)=\frac{1}{2}[\lambda_{i}(\Theta)+\lambda_{j}(\Theta)]\Sigma_{ij}^{(\alpha)}\nonumber
\end{eqnarray}
from which we obtain the entries of the symmetric logarithmic derivative~\cite{jing2014quantum} as
\begin{eqnarray}
\Sigma_{ij}^{(\alpha)}=
\frac{2\Phi_{ij}^{(\alpha)}(\Theta)}{\lambda_{i}(\Theta)+\lambda_{j}(\Theta)},
\end{eqnarray}
where
\begin{eqnarray}
\Phi_{ij}^{(\alpha)}(\Theta)=\frac{\partial}{\partial\theta_{\alpha}}\lambda_{i}(\Theta)\delta_{ij} + (\lambda_{j}-\lambda_{i})\mathbf{q}^{\top}_{i}(\Theta)\frac{\partial}{\partial\theta_{\alpha}}\mathbf{q}_{j}(\Theta)\nonumber
\end{eqnarray}
and $\delta_{ij}$ is the Kronecker function.

\bibliographystyle{apsrev4-1}
\bibliography{draft}

\end{document}